# SECULAR CYCLE OF THE NORTH-SOUTH SOLAR ASYMMETRY

K. GEORGIEVA, B. KIROV

*Solar-Terrestrial Influences Laboratory at the Bulgarian Academy of Sciences*

*Bl.3 Acad.G.Bonchev str., 1113 Sofia, Bulgaria*

**Abstract**

The North-South asymmetry of solar activity has been recognized for different solar phenomena. Following Waldmeier, it is now assumed that solar activity dominates in the Northern solar hemisphere during the ascending part of the secular solar cycle, in the Southern one during the descending part, and in epochs of secular minima and maxima the asymmetry is small. The episodes when this rule does not hold (the Maunder minimum in the end of the $17^{th}$ century, and $19^{th}$ and $20^{th}$ solar cycles in the $20^{th}$ century secular maximum) are considered as "anomalies". Analyzing solar activity influence on climate, we come to the conclusion that the asymmetry differs not in the ascending and descending parts of the secular solar cycles, but in consecutive secular cycles. This hypothesis is in agreement with all available data and leaves no anomalies.

## 1. Introduction

Different manifestations of solar activity are not identical in the two solar hemispheres. Waldmeier [1] based on data up to the $18^{th}$ solar cycle, found a close relation between the North-South asymmetry and the secular solar cycle, and formulated that solar activity dominates in the Northern solar hemisphere during the ascending part of the secular solar cycle, in the Southern one during the descending part, and in epochs of secular minima and maxima the asymmetry is small. However, there are at least two episodes when this rule does not hold: the secular maximum in the 20$^{th}$ century when the asymmetry had a maximum rather than a minimum, and end of the Maunder minimum in the end of the 17$^{th}$ century when the activity was clearly ascending and all sunspots were concentrates in the Southern solar hemisphere. In the present paper we find another relation between the long-term asymmetry and the secular solar cycle based on a study of the solar activity effects on climate.

**2. Solar activity and climate**

In spite of the great number of investigations performed so far and the extensive discussions [2-4], the problem of solar activity influences on climate on different time scales remains one of the most controversial ones in solar-terrestrial physics. For different sites and periods, positive, negative or missing correlations have been reported between surface air temperature and solar activity in the 11-year solar cycle [3, 4]. If the sign of the correlation changes randomly, this would be a serious reason to doubt the very existence of such a relation. On the other hand, if it shows some regularity, we could suppose there is some additional factor ruling the way in which solar activity affects climate.

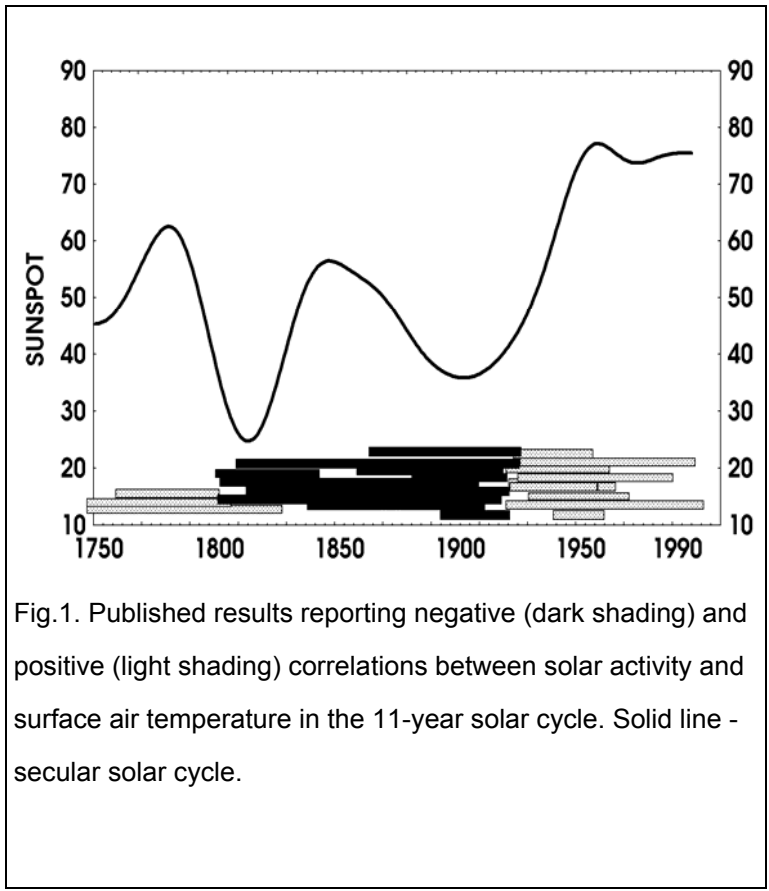

Fig.1. Published results reporting negative (dark shading) and positive (light shading) correlations between solar activity and surface air temperature in the 11-year solar cycle. Solid line - secular solar cycle.

As a first step, we make a compilation of the available published results - Fig.1. In the figure, the periods for which positive or negative correlations are reported between surface air temperature and solar activity in the 11-year cycle are presented by dark and light shading, respectively, and the secular solar cycle - by a solid line. It can be seen that the cases of positive and negative correlations are fairly well grouped, with the sign of the correlation depending on the period studied and not on the location, and changing in consecutive secular cycles.

This conclusion is derived from occasional, fragmentary observations. To verify it, we use three different reconstructions of global, hemispheric and zonal temperatures [5-7]. All of them yield similar results demonstrated in Fig.2. In the upper panel, the mean annual temperature (3 points running mean, detrended) is compared to the International Sunspot Number [8]. In the beginning of the period high temperatures are observed in solar minimum and low ones - at high solar activity. About 1920-30 the situation changes to the opposite. In the lower panel the cross-correlation functions for the two periods are presented.

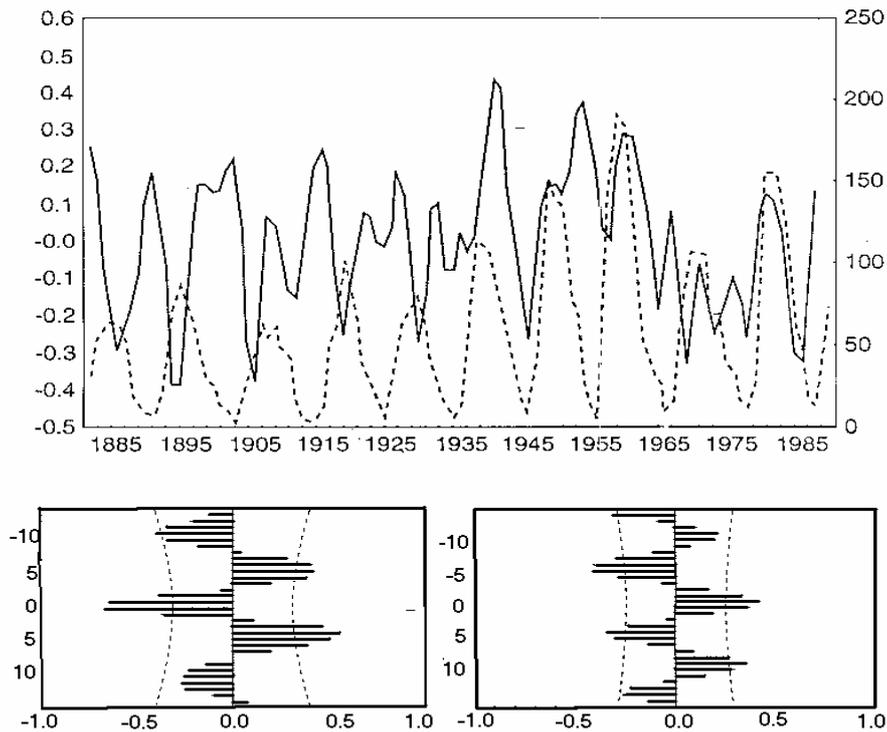

Fig.2. Upper panel: Mean zonal annual temperature (solid line) and sunspot numbers (dotted line); 5-point runnung mean, detrended; Lower panel: Cross-correlation functions for the first and second periods (see text)

All global reconstructions are relatively short, the longest data set [5] beginning in 1856. To study earlier epochs, data from individual meteorological stations with long measurement records have been used. In the Global Historical Climatology Network Temperature Database (GHCN) of NCDC [9] data are gathered from more than 8000 meteorological stations worldwide. Among them, 65 stations were found with almost continuous data records beginning before the middle of the 19[th] century. The time series for each station was divided into subseries yielding the best correlation with solar activity. In Fig.3 the relative number of the stations with statistically significant positive and negative correlation (along the positive and negative parts of the axis, respectively) is shown together with the centennial solar activity cycle. Fig.3 confirms that in the 18th century, in the vast majority of the available stations the surface air temperature was positively correlated with solar activity in the 11-year solar cycle, this correlation changed to negative in the 19th century and to positive again in the 20th century.

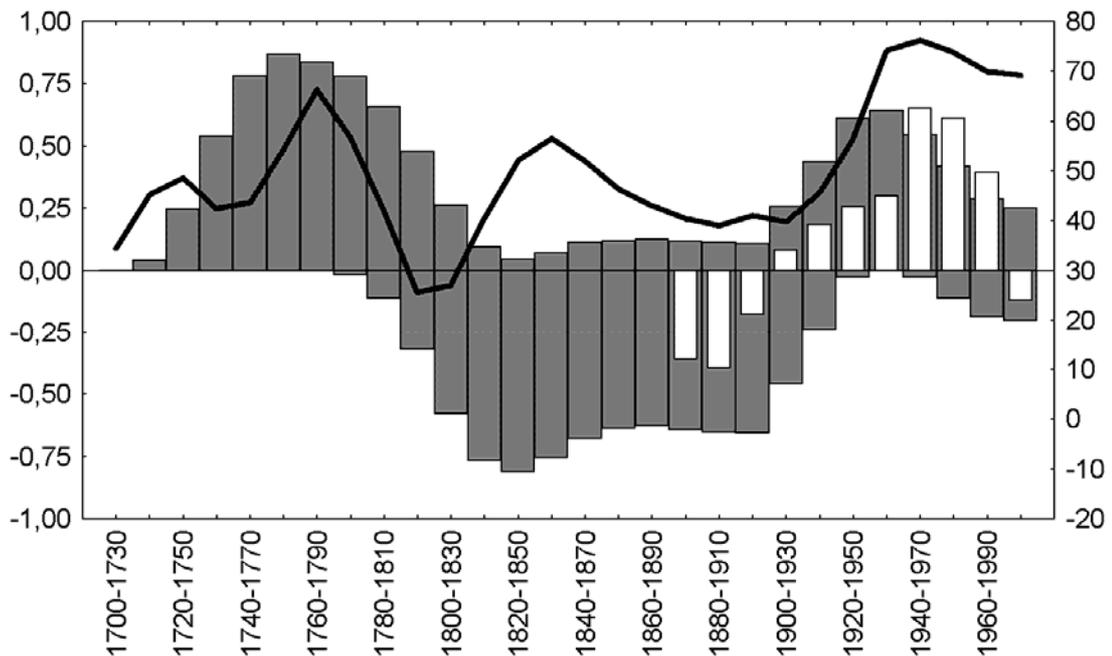

Fig.3. Dark bars - relative number of stations with statistically significant correlations between solar activity and surface air temperature in the 11-year solar cycle - positive or negative, along the positive and negative Y-axis, respectively; Solid line - secular solar cycle; White bars - North-South solar asymmetry.

### 3. Solar asymmetry and secular solar cycle

The next question is what solar activity parameters change in consecutive secular solar cycles. At present, no such parameter is known. So we try to approach the problem from the other side - the change of what solar activity parameters leads to a change in the sign of correlation between solar activity and surface air temperature. Smirnov [10, 11] has found that when the Earth passes from one sector of the interplanetary magnetic field into another, the sign of the correlation between the solar wind velocity and the atmospheric parameters changes. The sector boundary crossing means changing the solar hemisphere to which the Earth is exposed. So we could suppose that the activity originating from the two solar hemispheres has different effect upon atmospheric parameters. The white bars in Fig.3 represent the asymmetry of the solar activity, $A=(S_N-S_S)/(S_N+S_S)$ where $S_N$ and $S_S$ stand for the total sunspot area in the Northern and Southern solar hemispheres, respectively. Negative A seen in the 19$^{th}$ century means more active Southern solar hemisphere and is associated with negative correlation between solar activity and surface air temperature in the 11-year cycle, and positive A in the 20$^{th}$ century - more active Northern hemisphere and positive correlation.

We can therefore conclude that solar asymmetry is a parameter which changes in consecutive secular cycles, being positive in "even" cycles (if we denote the 20$^{th}$ century secular cycle as even) and negative in odd ones. It has a maximum positive or negative value coinciding with the maximum of solar activity in the secular cycle and changes sign around secular solar minimum. This leaves no "anomalies" in the available data. However, in order to be tested, additional data for solar asymmetry should be found from archive records for earlier periods. Solar asymmetry has officially been measured in Greenwich observatory only since 1874.

About 1970 the asymmetry becomes negative and part of the stations begin showing negative correlations between solar activity and surface air temperature in the 11-year cycle. If the above suggestion is correct, this would mean that the next secular cycle has already started. Yoshimura [12] and

Kopecky [13], using different methods, have both suggested that solar cycle 21 is already on the ascending branch of the next secular cycle.